\pdfoutput=1

\documentclass[twocolumn,showpacs]{revtex4}
\usepackage{amssymb,amsmath,graphicx}
\usepackage{color}
\usepackage[T1]{fontenc}
\begin{document}


\title{
  Lifshitz transition and thermoelectric properties of bilayer graphene
}

\author{Dominik Suszalski}
\affiliation{Marian Smoluchowski Institute of Physics, 
Jagiellonian University, {\L}ojasiewicza 11, PL--30348 Krak\'{o}w, Poland}
\author{Grzegorz Rut}
\affiliation{Marian Smoluchowski Institute of Physics, 
Jagiellonian University, {\L}ojasiewicza 11, PL--30348 Krak\'{o}w, Poland}
\author{Adam Rycerz}
\affiliation{Marian Smoluchowski Institute of Physics, 
Jagiellonian University, {\L}ojasiewicza 11, PL--30348 Krak\'{o}w, Poland}

\begin{abstract}
This is a~numerical study of thermoelectric properties of ballistic 
bilayer graphene in the presence of trigonal warping term in the
effective Hamiltonian. We find, in the
mesoscopic samples of the length $L>10\,\mu{}$m at sub--Kelvin temperatures, 
that both the Seebeck coefficient and the Lorentz number show anomalies 
(the additional maximum and minimum, respectively) when
the electrochemical potential is close to the Lifshitz energy, which
can be attributed to the presence of the van Hove singularity in a~bulk 
density of states. 
At higher temperatures the anomalies vanish, but measurable quantities 
characterizing remaining maximum of the Seebeck coefficient still unveil 
the presence of massless Dirac fermions
and make it possible to determine the trigonal warping strength. 
Behavior of the thermoelectric figure of merit ($ZT$) is also discussed. 
\end{abstract}

\date{February 15, 2018}
\maketitle

\section{Introduction}
It is known that thermoelectric phenomena provide a~valuable insight
into the details of electronic structure of graphene and other 
relativistic condensed-matter systems that cannot be solely determined
by conductance measurements \cite{Dol15}.  
Such a~fundamental perspective has inspired numerous studies on Seebeck 
and Nernst effects in mono- (MLG) and bilayer (BLG) graphenes 
\cite{Zue09,Wei09,Che09,Hwa09,Nam10,Wan10,Wan11,Wys13}
as well as in other two-dimensional systems \cite{New94,Cao15,Sex16,Sev17}. 
The exceptionally high thermal conductivity of graphenes has also 
drawn a~significant attention \cite{Pet11,Bal11,YXu14,Zha15,Cro16,Zha16} after 
a~seminal work by Balandin {\em et al.\/} \cite{Bal08}. 
A separate issue concerns thermal and thermoelectric properties of 
tailor-made graphene systems \cite{Dol15,Che10,Hua11,Lia12,Sev13,Hos15,Ann17}, 
including superlattices \cite{Che10}, 
nanoribbons \cite{Hua11,Lia12,Sev13,Hos15}, 
or defected graphenes \cite{Hos15,Ann17}, for which peculiar 
electronic structures may result in high thermoelectric figures 
of merit $ZT>2$ at room temperature \cite{Sev13,Hos15}. 

Unlike in conventional metals or semiconductors, thermoelectric power in 
graphenes can change a~sign upon varying the gate bias 
\cite{Zue09,Wei09,Che09}, making it
possible to design thermoelectronic devices that have no analogues in 
other materials \cite{Che15}.
In BLG the additional bandgap tunability \cite{Mac06b,Min07,Zha09} 
was utilized to noticeably enhance the thermoelectric power in 
a~dual-gated setup \cite{Wan11}. 

At sufficiently low temperatures, one can expect thermolectric properties 
of BLG to reflect most peculiar features of its electronic structure.
These features include the presence (in the gapless case)
of three additional Dirac points in the vicinity of each of primary Dirac 
points $K$ and $K'$ \cite{Mac06a,Orl12,Kat12,Mac13}. 
In turn, when varying chemical potential the system is expected to undergo 
the Lifshitz transition at $\mu=\pm{}E_L$ (the Lifshitz energy) \cite{Mac13}. 
What is more, electronic density of states (DOS) shows
van Hove singularities at $\mu=\pm{}E_L$. 
Unlike in systems with Mexican-hat band dispersion, for which diverging DOS
appears at the bottom of the conduction band and at the top of the valence 
band \cite{Sex16,Sev17}, in BLG each van Hove singularity separates 
populations of massless Dirac-Weyl quasiparticles ($|\mu|<E_L$) with 
approximately conical dispersion relation, and massive chiral quasiparticles 
($|\mu|>E_L$) characterized  by the effective mass 
$m_{\rm eff}\approx{}0.033\,m_e$, with $m_e$ being the free-electron mass. 
Although the value of $E_L$ is related to several
directly-measurable quantities, such as the minimal conductivity 
\cite{Mog09,Rut14b,Rut16}, available experimental results cover the full 
range of $E_L\sim{}0.1\!-\!1\,$meV \cite{Mac13}. 

\begin{figure}
\centerline{
  \includegraphics[width=\linewidth]{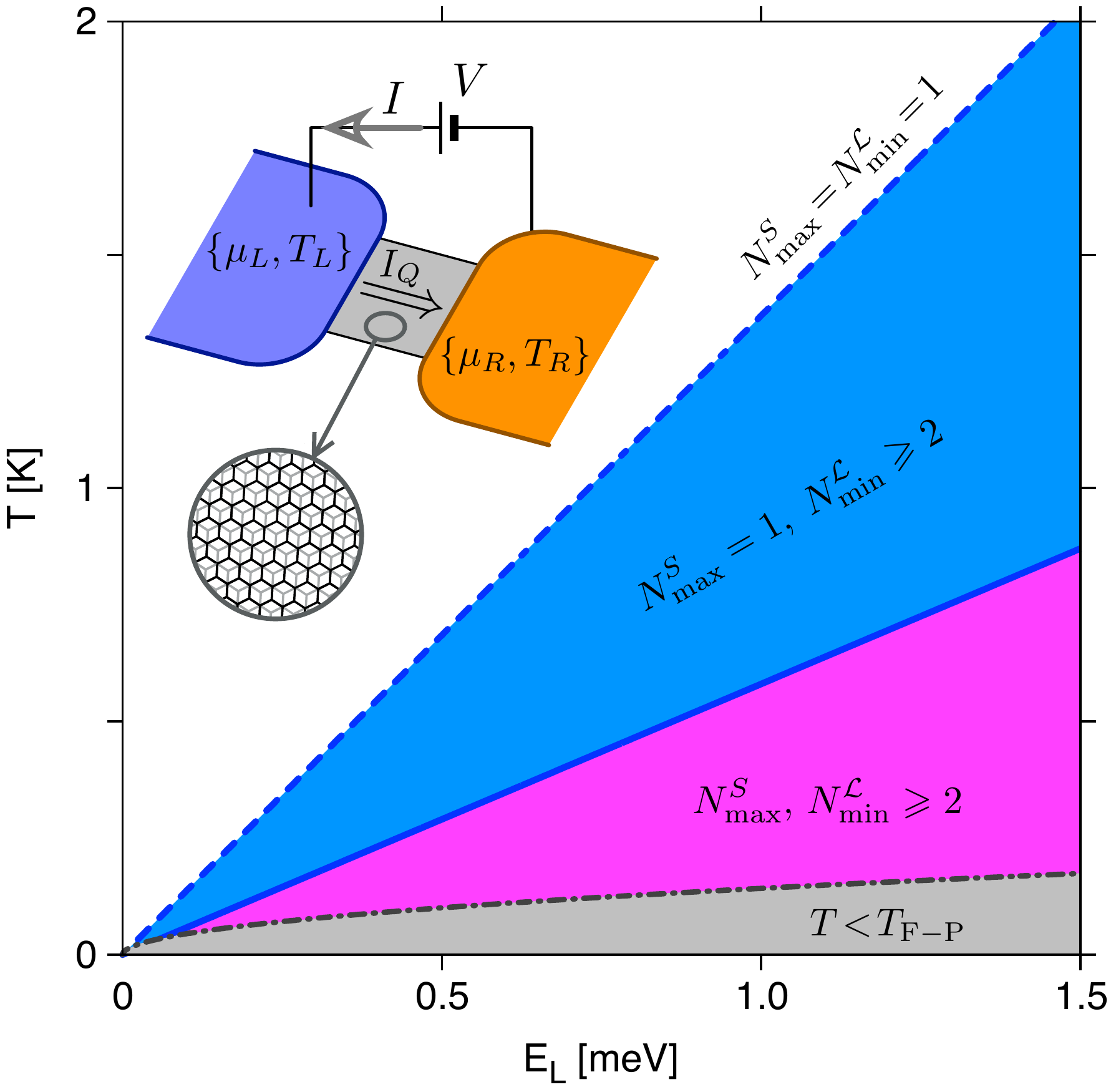}
}
\caption{ \label{phdiagfig} 
  The system studied numerically in the paper (inset) and the thermoelectric 
  phase diagram  (main) for Bernal-stacked bilayer graphene. 
  Thermal and electric currents ($I_Q,I$) flow between the leads, modeled 
  as infinitely-doped graphene regions with electrochemical potentials 
  $\mu_{L(R)}\rightarrow\infty$ (at a~fixed $\mu_{R}\!-\!\mu_{L}\equiv{}eV$, 
  with the electron charge $-e$ and the voltage $V$), 
  and temperatures $T_{L(R)}$, attached to the rectangular sample. 
  Additional gate electrodes (not shown) are used to tune the chemical
  potential in the sample area $\mu$ at zero bias between the layers. 
  The number of distinct maxima of the Seebeck coefficient ($N_{\rm max}^S$) 
  and minima of the Lorentz factor ($N_{\rm min}^{\cal L}$), occurring for 
  $0\!<\!\mu\!<\!\infty$, are indicated in the Lifshitz energy~--~temperature 
  parameter plane. The border of the Fabry-P\'{e}rot transport regime 
  $T\!<\!T_{\rm F-P}$ (in which $N_{\rm max}^S,N_{\rm min}^{\cal L}\gg{}1$), 
  corresponding to the system length $L=10^4\,l_\perp=17.7\,\mu$m,
  is also depicted. 
}
\end{figure}

The purpose of this work is to show that thermoelectric measurements in
ballistic BLG (see Fig.~\ref{phdiagfig}) can provide new insights into the 
nature of quasiparticles near the charge-neutrality point and allow one to 
estimate the Lifshitz energy. 
We consider a~relatively large, rectangular sample of ballistic 
BLG (with the length $L=17.7\,\mu$m, and the width $W=20\,L$), and
calculate its basic thermoelectric properties (including the Seebeck 
coefficient $S$ and the Lorentz number ${\cal L}$) within the 
Landauer-B\"{u}ttiker formalism \cite{Pau03,Esf06}. 
Our main findings are outlined in Fig.~\ref{phdiagfig}, where $N_{\rm max}^S$ 
($N_{\rm min}^{\cal L}$) -- the number of maxima (minima) of $S$ (${\cal L}$) 
appearing for $\mu>0$ is indicated in the
$E_L\,$--$\,T$ parameter plane. For instance, a~handbook value of 
$E_L/k_B\approx{}10\,$K \cite{Dre81} leads to the 
anomalies, including additional extrema at $\mu\approx{}E_L$, at sub--Kelvin 
temperatures. We further show that even for $T\gtrsim{}1\,$K (at which 
$N_{\rm max}^S=N_{\rm min}^{\cal L}=1$) the value of $E_L$ determines the carrier 
concentration corresponding to the remaining maximum of $S$ (or the minimum
of ${\cal L}$). 

The paper is organized as follows: The model and theory are described in Sec.~II, followed by the numerical results and discussions on the conductance, thermopower, validity of the Wiedemann-Franz law, the role of phononic thermal conductivity, and the figure of merit (Sec.~III). A~comparison with the linear model for transmission-energy dependence  (see Appendix~A) is also included. 
The conclusions are  given in Sec.~IV.

\section{Model and Theory}
\subsection{The Hamiltonian}
We start our analysis from the four-band effective Hamiltonian for low-energy excitations \cite{Mac13}, which can be written as
\begin{equation}
  \label{ham1val}
  H=\xi\left(\begin{array}{cccc}
      0 & v_F{}\pi & \xi\,t_{\perp} & 0\\
      v_F{}\pi^{\dagger} & 0 & 0 & v_3{}\pi\\
      \xi\,t_{\perp} & 0 & 0 & v_F{}\pi^{\dagger}\\
      0 & v_3{}\pi^{\dagger} & v_F{}\pi & 0
    \end{array}\right),
\end{equation}
where the valley index $\xi=1$ ($-1$) for $K$ ($K'$) valley, 
$v_F=\sqrt{3}\,t_0a/(2\hbar)\simeq{}10^6\,$m/s is the asymptotic Fermi 
velocity defined via the intralayer hopping $t_0=3.16\,$eV and the lattice 
parameter $a=0.246\,$nm, $\pi=\hbar{}e^{-i\theta}(-i\partial_x+\partial_y)$, 
$\theta$ denotes the angle between the main system axis and the armchair 
direction. 
(For the numerical calculations, we set $\hbar{}v_F=0.673\,$eV$\cdot$nm.)
The nearest-neighbor interlayer hopping is $t_\perp=0.381\,$eV \cite{Kuz09}
defining $l_\perp{}=\hbar{}v_F/t_\perp=1.77\,$nm, 
and $v_3=v_F{}t'/t_{0}$ with $t'$ being the next-nearest neighbor 
(or {\em skew}) interlayer hopping. 

The Hamiltonian (\ref{ham1val}) leads to the bulk dispersion relation for 
electrons \cite{Mac06a,Mac13} 
\begin{align}
  E_\pm^{(e)}(\mbox{\boldmath $k$}) &= 
  \left[
    \frac{1}{2}t_\perp^2+\left(v_F^2+\frac{1}{2}v_3^2\right)k^2
    \pm\sqrt{\Gamma(\mbox{\boldmath $k$})}
  \right]^{1/2},  \nonumber \\
  \Gamma(\mbox{\boldmath $k$}) &= 
  \frac{1}{4}\left(t_\perp^2\!-\!\hbar^2v_3^2k^2\right)^2+
  \hbar^2v_F^2k^2\left(t_\perp^2\!+\!\hbar^2v_3^2k^2\right) \label{enkk} \\
  &+ 2\xi{}\,t_\perp{}\hbar^3v_3v_F^2k^3\cos{}3\varphi, \nonumber
\end{align}
where $\mbox{\boldmath $k$}\equiv{}(k_x,k_y)$ is the in-plane wavevector
(with $\mbox{\boldmath $k$}=0$ referring to K or K' point), 
$k=|\mbox{\boldmath $k$}|$, and the angle $0\leqslant\varphi<2\pi$ can be 
defined as the argument $\mbox{arg}\,z$ of a~complex number 
\begin{equation}
z=e^{-i\theta}(k_x+ik_y). 
\end{equation}
For holes, we have 
$E_\pm^{(h)}(\mbox{\boldmath $k$})=-E_\pm^{(e)}(\mbox{\boldmath $k$})$
\cite{elhofoo}. 

\begin{figure}
\centerline{
  \includegraphics[width=\linewidth]{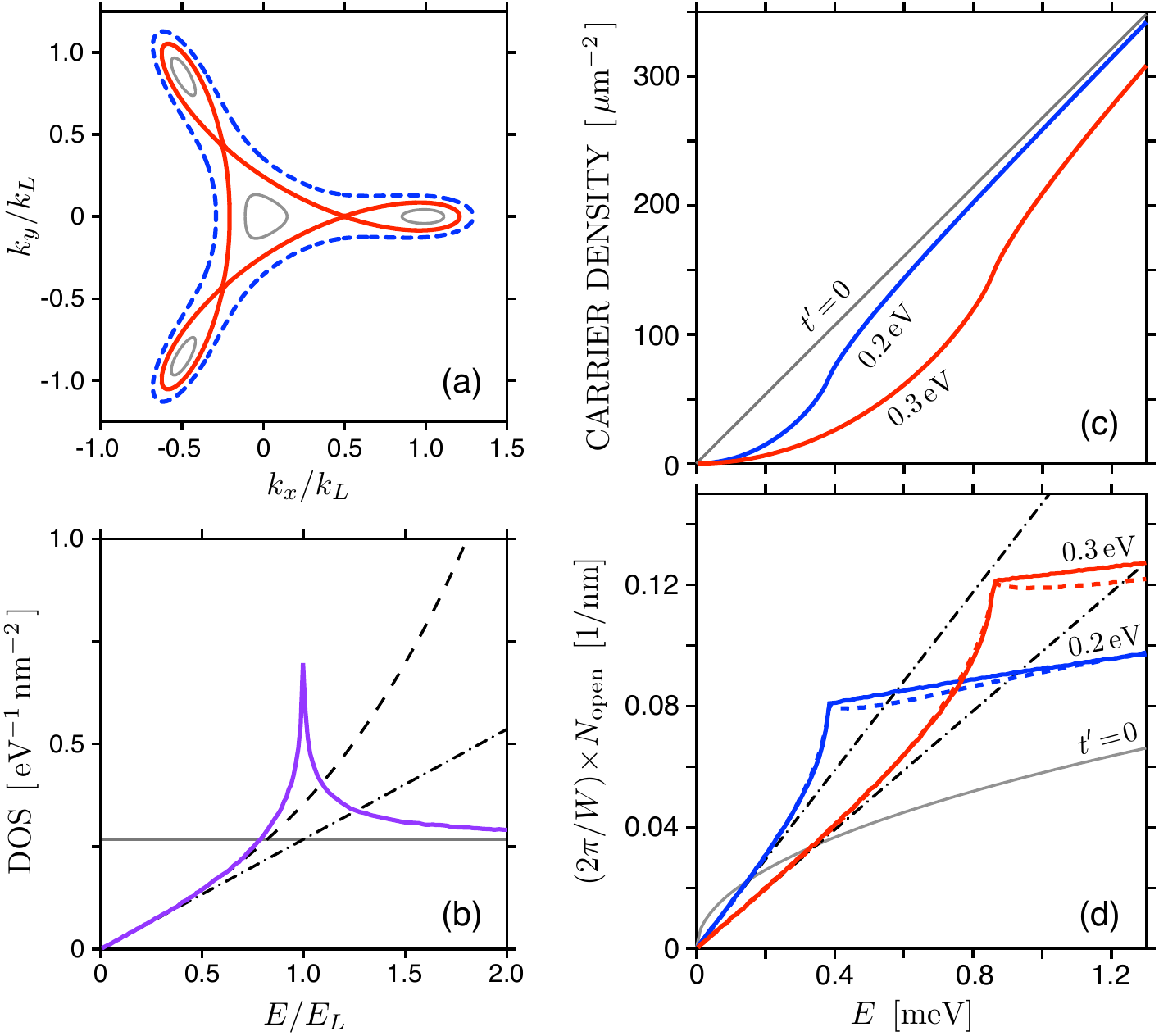}
}
\caption{ \label{endonopfig}
  Physical consequences of the dispersion relation given by 
  Eq.\ (\ref{enkk}). 
  (a) Equienergy surfaces for $E=0.5\,E_L$ (grey solid lines), 
  $E=E_L$ (red solid line), and $E=1.5\,E_L$ (blue dashed line)
  for the crystallographic orientation $\theta=0$. 
  (b) Density of states (purple solid line) and the approximating 
  expressions given by Eqs.\ (\ref{rhozero}), (\ref{rhoapp}), 
  and (\ref{rholin}) (grey solid, black dashed, and black dotted-dashed 
  line, respectively). 
  (c) Carrier density and (d) the number of open channels for different
  values of $t'$ (specified for each line). Solid and dashed lines in panel 
  (d) corresponds to $\theta=0$ and $\theta=\pi/6$, dotted-dashed lines
  represent the approximating Eq.\ (\ref{loklin}). 
}
\end{figure}

\subsection{Low-energy electronic structure}
Basic consequences of Eq.\ (\ref{enkk}) are illustrated 
in Fig.\ \ref{endonopfig}. 
In the energy range $|E|<E_L$, with the Lifshitz energy
\begin{equation}
  \label{elifdef}
  E_L=\frac{1}{4}t_\perp\left(\frac{v_3}{v_F}\right)^2, 
\end{equation}
there are four distinct parts of the Fermi surface (see 
Fig.\ \ref{endonopfig}a), centered at $z=z_0,\dots,z_3$, where $z_0=0$, 
$z_j=k_L\exp(2\pi{}ij/3), j=1,2,3$, and 
\begin{equation}
  \label{klif}
  k_L = \frac{t_\perp{}v_3}{\hbar{}v_F^2}. 
\end{equation}
For $|E|\geqslant{}E_L$ the Fermi surface becomes connected, and the 
transition at $E=\pm{}E_L$ is accompanied by the van Hove singularity in the 
density of states $\rho(E)$ (see Fig.\ \ref{endonopfig}b), 
which can be defined (for electrons) via 
\begin{equation}
  \int_0^EdE'\rho({E'}) \equiv n(E) = \frac{{\cal A}(E)}{\pi^2},
\end{equation}
where $n(E)$ is the physical carrier density (taking into account spin
and valley degeneracies $g_s=g_v=2$) depicted in Fig.\ \ref{endonopfig}c, 
and ${\cal A}(E)$ denotes the area bounded by the Fermi surface in the 
$(k_x,k_y)$ plane \cite{mcycfoo}. 
In particular, taking the limit of $t'\rightarrow{}0$ we have
\begin{equation}
\label{rhozero}
  \rho_{\,t'\rightarrow{}0}(|E|\ll{}t_\perp)\approx
  \frac{2m_{\rm eff}}{\pi{}\hbar^2}= \frac{t_\perp}{\pi{}(\hbar{}v_F)^2}
  \equiv{}\rho_0,
\end{equation}
where we have introduced the effective mass relevant in the absence of
trigonal warping ($E_L=0$). 
At finite $t'$ ($E_L>0$) the value of $\rho_0$ defined in Eq.\ (\ref{rhozero}) 
is approached by the actual $\rho(E)$ for $|E|\gtrsim{}E_L$
(see Fig.\ \ref{endonopfig}b). 
Also, in the $t'\neq{}0$ case, we find that the approximating formula 
\begin{equation}
  \label{rhoapp}
  \rho(E) \approx 
  \frac{\rho_0|E|}{E_L}\left[1+0.33\left({E}/{E_L}\right)^2\right],
\end{equation}
reproduces the actual $\rho(E)$ with $1\%$ accuracy for $|E|\leqslant{}E_L/2$ 
(being the energy interval most relevant for discussion presented in the 
remaining parts of this paper). 
For $|E|\ll{}E_L$, leaving only the leading term on the right-hand side of 
Eq.\ (\ref{rhoapp}) brought us to
\begin{equation}
  \label{rholin}
  \rho(|E|\ll{}E_L)\approx 
  \frac{4|E|}{\pi(\hbar{}v_3)^2}, 
\end{equation}
which can be interpreted as a double-monolayer DOS with the Fermi velocity 
replaced by $v_3$. 

Although the trigonal-warping effects become hardly visible in $\rho(E)$ for
$E\gtrsim{}E_L$, characteristic deformations of the Fermi surface can be
noticed also for $E\gg{}E_L$. We point out that a~compact quantity taking 
this fact into account, which can be determined directly from 
Eq.\ (\ref{enkk}) without resorting to quantum transport simulations,
is the number of propagating modes ({\em open channels}) 
$N_{\rm open}(\theta,E)$ presented in Fig.\ \ref{endonopfig}d. 
It can be defined as a~total number of solutions, with real $k_x$, 
of equations 
\begin{equation} 
  E^{(p)}_+(k_x,q\Delta{}k_y)=E, \ \ \ \ \ 
  E^{(p)}_-(k_x,q\Delta{}k_y)=E, 
\end{equation}
where $p=e$ for electrons ($E>0$) or $p=h$ for holes ($E<0$), 
$q=0,\pm{}1,\pm{}2,\dots$, and $\Delta{}k_y=2\pi/W$ (we suppose 
the periodic boundary conditions along the $y$-axis), 
that correspond to a~chosen sign the group velocity, e.g.\  
$\left(v_g\right)^{(p)}_{\pm,q}=
\partial{}E^{(p)}_\pm(k_x,q\Delta{}k_y)/\partial{k_y}>0$. 
Apart from the $t'\rightarrow{}0$ limit, for which 
\begin{equation}
  \label{loktp0}
  N_{\rm open}(t'\rightarrow{}0,|E|\ll{}t_\perp) \approx
  2\,\sqrt{\frac{{|E|t_\perp}}{(\hbar{}v_F)^2}}\,\frac{1}{\Delta{}k_y}, 
\end{equation}
the number of open channels is anisotropic and shows the periodicity with 
a~period $\pi/3$. In the low-energy limit 
\begin{equation}
  \label{loklin}
  N_{\rm open}(\,\theta,|E|\ll{}E_L\,)  \approx 
  2F(\theta) \,\frac{|E|}{\hbar{}v_3}\,\frac{1}{\Delta{}k_y},
\end{equation}
where
\begin{align}
  F(\theta) &= 1+\sum_{j=1,2,3}\left[
    1-\frac{8}{9}\cos^2\left(\theta+\frac{2\pi}{3}j\right)\right]^{1/2},  \\
  & \approx 3.126+0.029\,\cos{}6\theta. \nonumber
\end{align}
The anisotropy is even more apparent for $|E|\gtrsim{}E_L$. 
In particular, $N_{\rm open}(\theta=0,E)$ grows monotonically with increasing 
$E$, whereas $N_{\rm open}(\theta=\pi/6,E)$ has a~shallow minimum at 
$E\approx{}1.11{}E_L$. 

It is also visible in Fig.\ \ref{endonopfig}d that the effects of increasing 
$t'$ are essentially opposite at different energy ranges: 
For $|E|\gtrsim{}E_L$, $N_{\rm open}$ grows systematically with $t'$; 
for $|E|\ll{}E_L$ we have $N_{\rm open}\propto{}1/t'$ following from 
Eq.\ (\ref{loklin}). Such a~feature has no analogues in behaviors 
of other characteristics presented in Fig.\ \ref{endonopfig}.

\subsection{Thermoelectric properties}
In the linear-response regime, thermoelectric properties of a~generic
nanosystem in graphene are determined via Landauer-B\"{u}ttiker expressions 
for the electrical and thermal currents \cite{Lan57,But85}
\begin{align}
  I &= -\frac{g_sg_ve}{h}\int{}dE\,T(E)\left[f_L(E)\!-\!f_R(E)\right],  \\
  I_Q &= \frac{g_sg_v}{h}\int{}dE\,T(E)\left[f_L(E)\!-\!f_R(E)\right] 
  (E\!-\!\mu), \label{iqland}
\end{align}
where $g_s=g_v=2$ are spin and valley degeneracies, 
$T(E)\equiv{}\mbox{Tr}({\bf t}{\bf t}^\dagger)$ with ${\bf t}$ being
the transmission matrix \cite{Rut14b}, $f_{L(R)}$ is the distribution 
functions for the left (right) lead with electrochemical potential 
$\mu_{L(R)}$ and temperature $T_{L(R)}$. 
Assuming that $\mu_L-\mu_R\equiv{}-eV$ and $T_L-T_R\equiv{}\Delta{}T$ are 
infinitesimally small [hereinafter, we refer to the averages 
$\mu=(\mu_L+\mu_R)/2$ and $T=(T_L+T_R)/2\,$], we obtain the conductance $G$, 
the Seebeck coefficient $S$, and the electronic part of the thermal 
conductance $K_{\rm el}$, as follows \cite{Esf06}
\begin{align}
  G &= \left.\frac{I}{V}\right|_{\Delta{}T=0} = e^2L_0, 
  \label{gland} \\
  S &= -\left.\frac{V}{\Delta{}T}\right|_{I=0} = \frac{L_1}{eTL_0}, 
  \label{sland} \\
  K_{\rm el} &= \left.\frac{I_Q}{\Delta{}T}\right|_{I=0} = 
  \frac{L_0L_2-L_1^2}{TL_0}, 
  \label{keland}
\end{align}
where $L_n$ (with $n=0,1,2$) is given by
\begin{equation}
  \label{llndef}
  L_n=\frac{g_sg_v}{h}\int{}dE\,T(E)\left(
    -\frac{\partial{}f_{\rm FD}}{\partial{}E}\right)(E-\mu)^n, 
\end{equation}
with $f_{\rm FD}(\mu,T,E)=
1/\left[\,\exp\left((E\!-\!\mu)/k_BT\right)+1\,\right]$ 
the Fermi-Dirac distribution function. 

By definition, the Lorentz number accounts only the electronic part of 
the thermal conductance, 
\begin{equation}
  \label{lodef}
  {\cal L}=\frac{K_{\rm el}}{TG}=
  \frac{L_0L_2-L_1^2}{e^2T^2L_0^2}.  
\end{equation}
The thermoelectric figure of merit accounts the total thermal conductance
($K_{\rm tot}=K_{\rm el}+K_{\rm ph}$)
\begin{equation}
  \label{ztdef}
  ZT=\frac{TGS^2}{K_{\rm tot}}=\left(\frac{K_{\rm el}}{K_{\rm el}+K_{\rm ph}}\right)
  \frac{L_1^2}{L_0L_2-L_1^2}, 
\end{equation}
where the phononic part can be calculated using 
\begin{equation}
\label{kaphland}
  K_{\rm ph}=\frac{1}{2\pi}\int{}d\omega\,\hbar{}\omega
  \frac{\partial{}f_{\rm BE}}{\partial{}T}{\cal T}_{\rm ph}(\omega), 
\end{equation}
with $f_{\rm BE}(T,\omega)=
1/\left[\,\exp\left(\hbar\omega/k_B{}T\right)-1\,\right]$
the Bose-Einstein distribution function and ${\cal T}_{\rm ph}(\omega)$
the phononic transmission spectrum. 
For BLG in a~gapless case considered in this work, we typically have 
$K_{\rm ph}\sim{}K_{\rm el}$ (see Sec.~IIID) \cite{phonofoo}. 
As ${\cal T}_{\rm ph}(\omega)$ in Eq.\ (\ref{kaphland}) is generally much
less sensitive to external electrostatic fields than $T(E)$ in 
Eq.\ (\ref{keland}) it should be possible --- at least in principle --- 
to independently determine $K_{\rm ph}$ and $K_{\rm el}$ in the experiment. 

It can be noticed that ultraclean ballistic graphene shows approximately
linear transmission to Fermi-energy dependence $T(E)\propto{}|E|$ (where
$E=0$ corresponds to the charge-neutrality point) 
\cite{Two06,Dan08,Kum16,limofoo}. 
Straightforward analysis (see Appendix~A) leads to extremal values of 
the Seebeck coefficient as a~function of the chemical potential
\begin{equation}
\label{smaxlimo}
S_{\rm max}=-S_{\rm min}\approx{}k_B/e=86\,\mu\text{V/K}, 
\end{equation}
providing yet another example of a~material characteristic given solely
by fundamental constants \cite{Two06,Nai08}. 
Similarly, the Lorentz number reaches, at $\mu=0$, the maximal value
given by
\begin{equation}
\label{lormaxlimo}
{\cal L}_{\rm max}=
  \frac{9\,\zeta(3)}{2\ln{}2}\left(\frac{k_B}{e}\right)^2 = 
2.37\,{\cal L}_{\rm WF}, 
\end{equation}
with ${\cal L}_{\rm WF}=\frac{1}{3}\pi^2(k_B/e)^2$ being the 
familiar Wiedemann-Franz constant. Although the disorder and electron-phonon
coupling may affect the above-mentioned values, existing experimental works 
report $S_{\rm max}$ and ${\cal L}_{\rm max}$ close to the given by Eqs.\
(\ref{smaxlimo}) and (\ref{lormaxlimo}) for both MLG and BLG, provided the 
temperature is not too low \cite{Zue09,Wei09,Che09,Hwa09,Nam10,Wan10,Cro16}. 

At low temperatures, the linear model no longer applies, partly due 
to the contribution from evanescent modes \cite{Two06,Sny07}, and partly due
to direct trigonal-warping effects on the electronic structure (see
subsection IIB). 
For this reason, thermoelectric properties calculated numerically 
from Eqs.\ (\ref{gland})--(\ref{ztdef}) are discussed next.

\section{Results and discussion}

\subsection{Zero-temperature conductivity}

For $T\rightarrow{}0$ Eq.\ (\ref{gland}) leads to the conductivity
\begin{equation}
\label{glandzero}
  \sigma(T\!\rightarrow{}\!0) = \frac{G(T\!\rightarrow{}\!0)L}{W}=
  \frac{g_0L}{W}\,\mbox{Tr}({\bf t}{\bf t}^\dagger), 
\end{equation}
with the conductance quantum $g_0=4e^2/h$. 
As the right-hand side of Eq.\ (\ref{glandzero}) is equal to 
$\mbox{Tr}({\bf t}{\bf t}^\dagger)$ with a~constant prefactor, 
$\sigma(T\!\rightarrow{}\!0)$ gives a~direct insight into the 
transmission-energy dependence that defines all the thermoelectric properties 
[see Eqs.\ (\ref{gland})--(\ref{ztdef})]. 

In order to determine the transmission matrix ${\bf t}$ for a~given 
electrochemical potential $\mu$ we employ the computational scheme similar 
to the presented in Ref.\ \cite{Rut14b}. 
However, at finite-precision arithmetics, the mode-matching equations 
become ill-defined for sufficiently large $L$ and $\mu$, as they contain
both exponentially growing and exponentially decaying coefficients. 
This difficulty can be overcome by dividing the sample area into $N_{\rm div}$
consecutive, equally-long parts, and matching wave functions for all 
$N_{\rm div}\!+\!1$ interfaces \cite{ndivfoo}.

\begin{figure}
\centerline{
  \includegraphics[width=\linewidth]{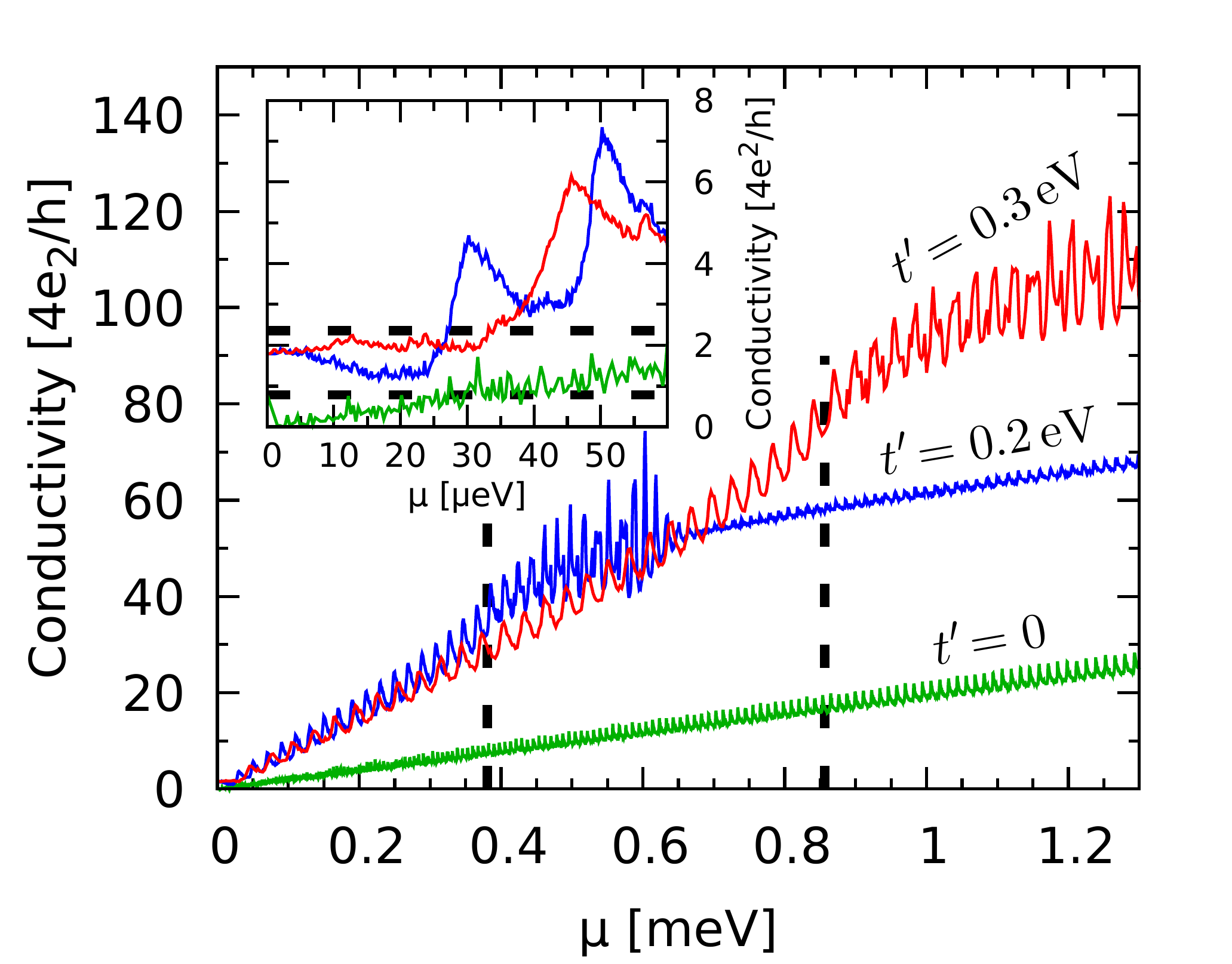}
}
\caption{ \label{sigte0fig}
  Zero-temperature conductivity [see Eq.\ (\ref{glandzero}) in 
  the main text] for $L=W/20=10^4\,l_\perp=17.7\,\mu$m and $\theta=0$ 
  \cite{theta0foo} plotted as a~function of the chemical potential. 
  The value of skew-interlayer hopping $t'$ is specified for each line. 
  Remaining tight-binding parameters are given below Eq.\ (\ref{ham1val})
  in the main text. Vertical lines mark values of the Lifshitz energy, 
  given by Eq.\ (\ref{elifdef}), for $t'=0.2\,$eV and $t'=0.3\,$eV. 
  Inset is a~zoom in, for low chemical potentials, with horizontal lines
  depicting $\sigma=2\,\sigma_{\rm MLG}=(8/\pi)\,e^2/h$ and 
  $\sigma=6\,\sigma_{\rm MLG}$. 
}
\end{figure}

Numerical results are presented in Fig.\ \ref{sigte0fig}. 
A~striking feature of all datasets is the presence of quasiperiodic 
oscillations of the Fabry-P\'{e}rot type. Although such oscillations can be
regarded as artifacts originating from a~perfect, rectangular shape of the
sample area (vanishing immediately when e.g.\ samples with nonparallel edges
are considered, see Ref.\ \cite{Rut14c}) their periodic features are useful 
to benchmark the numerical procedure applied. 

In particular, for $t'=0$, the conductivity shows abrupt 
features at energies associated with resonances at normal incidence 
($k_y=0$) \cite{Sny07}, namely 
\begin{equation}
  \label{enrtp0}
  E_n(t'=0)\approx{} \pm{}\hbar{}v_F{}l_\perp \left(\frac{\pi{}n}{L}\right)^2,
  \ \ \ \ 
  n=1,2,3,\dots,
\end{equation}
where the approximation refers to the parabolic dispersion relation applying
for $|E_n|\ll{}t_\perp$, or equivalently for 
$n\ll{}L/(\pi{}l_\perp)\approx{}3180$ in our numerical example. 
In turn, the separation between consecutive resonances is
\begin{align}
  \Delta{}E_n(t'=0) &= |E_{n+1}\!-\!E_{n}|\approx{} 
  \frac{2n\!+\!1}{t_\perp}\left(\frac{\pi{}\hbar{}v_F}{L}\right)^2 
  \nonumber\\
  &\approx 2\,\frac{\pi{}\hbar{}v_F}{L}\sqrt{\frac{|E_n|}{t_\perp}},  
  \label{denrtp0}
\end{align}
with the last approximation corresponding to $n\gg{}1$. 

For $t'\neq{}0$ the analysis is much more cumbersome even at low energies, 
as we have resonances associated with four distinct Dirac cones. 
However, resonances at normal incidence associated with the central cone, 
occurring at $E_n\approx{}\pi\hbar{}v_3n/L$ ($n=\pm{}1,\pm{}2,\dots$), 
allow us to estimate the order of magnitude of the relevant separation as
\begin{align}
  \Delta{}E_n(t'\neq{}0) &\sim \frac{\pi\hbar{}v_3}{L} =
  2\,\frac{\pi{}\hbar{}v_F}{L}\sqrt{\frac{E_L}{t_\perp}}
  \equiv{}k_BT_{\rm F-P}, 
  \label{denrtpp}
\end{align}
finding that the period of Fabry-P\'{e}rot oscillations is now 
energy-independent and should be comparable with $\Delta{}E_n(0)$ 
given by Eq.\ (\ref{denrtp0}) for $\mu=E_n\approx{}E_L(t')$. 
The data displayed in Fig.\ \ref{sigte0fig} show that the oscillation period 
is actually energy-independent in surprisingly-wide interval of 
$\mu\lesssim{}1.5E_L$, with the multiplicative factor  
$\left.\Delta{}E_n(t')/\Delta{}E_n(0)\right|_{\mu=E_L(t')}\approx{}3$.  
The oscillation amplitude is also enhanced, in comparison to the $t'=0$ 
case, for $\mu\lesssim{}1.5E_L$. 
For $\mu\gtrsim{}1.5E_L$, both the oscillation period and amplitude are 
noticeably reduced, resembling the oscillation pattern observed for 
the $t'=0$ case. 
It is also visible in Fig.\ \ref{sigte0fig}, that the mean conductivity
(averaged over the oscillation period) linearly increase with $\mu$ for 
$\mu\lesssim{}E_L$, with a~slope weakly dependent on $t'$. 
Such a~behavior indicates that $\mbox{Tr}({\bf t}{\bf t}^\dagger) <
N_{\rm open}$ [see Fig.\ \ref{endonopfig}(d)] what can be interpreted as
a~backscattering (or transmission reduction) appearing when different classes 
of quasiparticles are present in the leads and in the sample area. 
For larger $\mu$, the transmission reduction is still significant, but 
its dependence on $t'$ is weaken, and the sequence of lines from  
Fig.\ \ref{endonopfig}(d) is reproduced. 

Detailed explanation of the above-reported observations, in terms of 
simplified models relevant for $|\mu|\ll{}E_L$ and for $|\mu|\gtrsim{}E_L$, 
will be presented elsewhere. Here we only notice that the linear model 
for transmission-energy dependence is justified, for $|\mu|\lesssim{}E_L$, 
with the numerical results presented in Fig.\ \ref{sigte0fig}.

The rightmost equality in Eq.\ (\ref{denrtpp}) defines the Fabry-P\'{e}rot
temperature, which can be written as
\begin{equation}
  T_{\rm F-P}=\frac{\pi{}\,t_\perp{}t'l_\perp}{k_B{}t_0{}L}=
  13890\,\text{K}\times{}\frac{t'\,l_\perp}{t_0L}. 
\end{equation}
For $L=10^4{}\,l_\perp$, we obtain $T_{\rm F-P}=88\,$mK if $t'=0.2\,$eV, or 
$T_{\rm F-P}=132\,$mK if $t'=0.3\,$eV. 
For higher temperatures, Fabry-P\'{e}rot oscillations are smeared out due to
thermal excitations involving transmission processes from a~wider energy 
window [see Eqs.\ (\ref{gland}) and (\ref{llndef})].

\subsection{Thermopower and Wiedemann-Franz law}

\begin{figure}
\centerline{
  \includegraphics[width=0.9\linewidth]{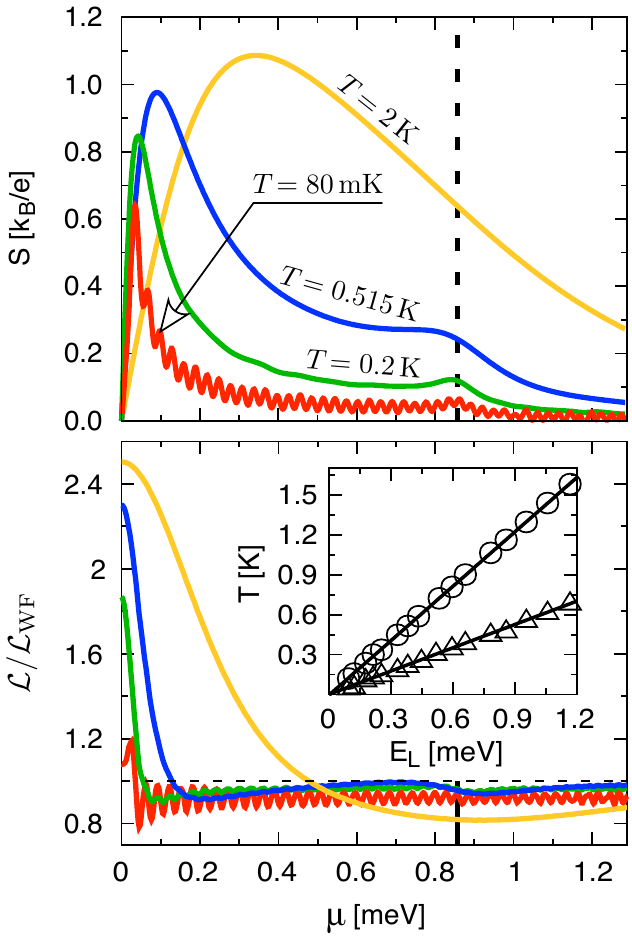}
}
\caption{ \label{seebwffig}
  Seebeck coefficients $S$ and Lorentz number ${\cal L}$ for $t'=0.3\,$eV,
  as a~function of the chemical potential.  The temperature is specified for 
  each line in top panel and is the same in both panels. 
  Vertical lines mark the Lifshitz energy; horizontal line in bottom panel 
  corresponds to the Wiedemann-Franz value ${\cal L}={\cal L}_{\rm WF}=
  \frac{1}{3}\pi^2(k_B/e)^2$.  
  Inset shows crossover temperatures, corresponding to vanishing of the 
  secondary maximum of $S$ (triangles) and minimum of ${\cal L}$ (circles), 
  plotted as functions of the Lifshitz energy, together with the best-fitted 
  linear functions [see Eqs.\ (\ref{tseebfit}) and (\ref{tlorefit})]. 
}
\end{figure}

As the finite-$T$ conductivity is simply given by a~convolution of 
$T(E)=\mbox{Tr}({\bf t}{\bf t}^\dagger)$ with the derivative of the Fermi-Dirac
function, we proceed directly to the numerical analysis of the Seebeck 
coefficient and the Lorentz number given by 
Eqs.\ (\ref{sland})--(\ref{lodef}) \cite{mottfoo}. 
In Fig.\ \ref{seebwffig}, these thermoelectric properties are displayed 
as functions of $\mu$, for a~fixed $t'=0.3\,$eV (corresponding to 
$E_L/k_B\approx{}10\,$K) and varying temperature. 
Quasiperiodic oscillations are still prominent in datasets for the lowest 
presented temperature, $T=80\,\text{mK}\approx{}0.6\,T_{\rm F-P}$, although 
it is rather close to $T_{\rm F-P}$. 
This is because all the abrupt features of $T(E)$ are magnified when 
calculating $S$, or ${\cal L}$, since they affect the nominator and 
the denominator in the corresponding Eq.\ (\ref{sland}), or 
Eq.\ (\ref{lodef}), in a~different manner. 
For $T=0.2\,\text{K}\approx{}1.5\,T_{\rm F-P}$ the oscillations vanish for $S$
and are strongly suppressed for ${\cal L}$; 
instead, we observe the anomalies: the secondary maximum of $S$ and minimum 
of ${\cal L}$, located near $\mu=E_L$. 
The secondary maximum of $S$ vanishes for $T=T_{\star}^S=0.515\,$K, but
${\cal L}$ still shows the two shallow minima at this temperature. 
(We find that the minima of ${\cal L}$ merge at $T_{\star}^{\cal L}=1.20\,\text{K}
=2.33\,T_{\star}^S$, the corresponding dataset is omitted for clarity.) 
For $T=2\,K$, each of $S$ and ${\cal L}$ shows a~single extremum for $\mu>0$. 

The crossover temperatures $T_{\star}^S$ and $T_{\star}^{\cal L}$ as functions of 
$E_L$, varied in the range corresponding to $0.1\,\text{eV}\leqslant{}t'
\leqslant{}0.35\,\text{eV}$, are also plotted in Fig.\ \ref{seebwffig} 
(see the inset). 
The least-squares fitted lines are given by
\begin{align}
  T_{\star,{\rm fit}}^S = 0.0504(5) \times E_L/k_B, \label{tseebfit} \\
  T_{\star,{\rm fit}}^{\cal L} = 0.1176(3) \times E_L/k_B, \label{tlorefit}
\end{align}
with standard deviations of the last digit specified by numbers in 
parentheses. 

These findings can be rationalized by referring to the onset on low-energy 
characteristics given in Sec.\ IIB (see Fig.\ \ref{endonopfig}). 
In particular, the abrupt features of $T(E)$ near $E=E_L$, attributed 
to the van Hove singularity of $\rho(E)$ shown in Fig.\ \ref{endonopfig}(b), 
or to the anisotropy of $N_{\rm open}(\theta,E)$ in Fig.\ \ref{endonopfig}(d), 
are smeared out when calculating thermoelectric properties for
energies of thermal excitations
\begin{equation}
  k_BT\gtrsim{}0.1\,E_L. 
\end{equation}
However, some other features, related to trigonal-warping effects on  
$N_{\rm open}(\theta,E)$ or $n(E)$ [see Fig.\ \ref{endonopfig}(c)] away from 
$E=E_L$, visible in thermoelectric properties, may 
even be observable at higher temperatures. 

\subsection{Comparison with the linear model for transmission-energy 
dependence}

\begin{figure}
\centerline{
  \includegraphics[width=0.9\linewidth]{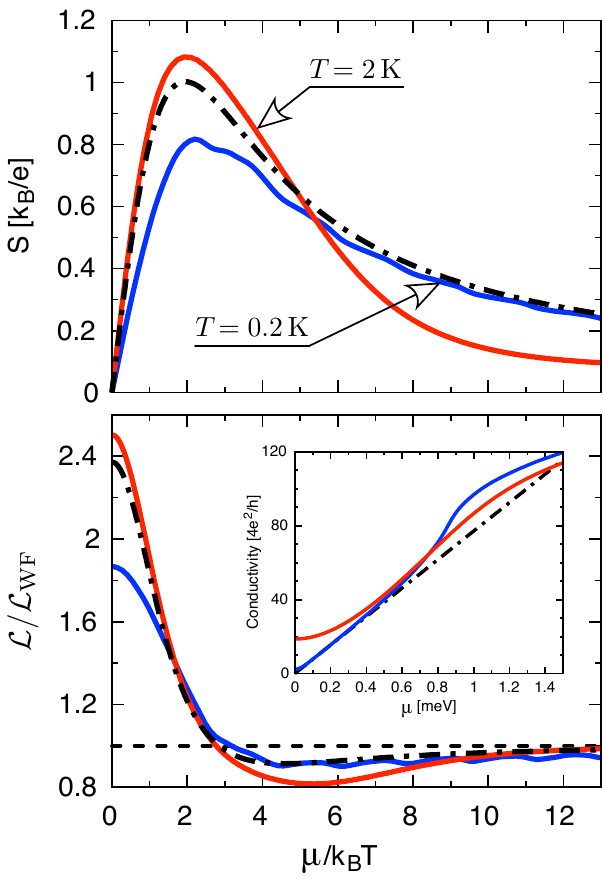}
}
\caption{ \label{polylofig}
  Same as Fig.\ \ref{seebwffig} but plotted versus the dimensionless
  variable $\mu/(k_BT)$. Solid lines in both panels represent the datasets 
  for $T=0.2\,$K and $T=2\,$K. Dashed-dotted lines correspond to 
  Eqs.\ (\ref{seebvials}) and (\ref{lorevials}) in Appendix~A, following from 
  the linear model for transmission-energy dependence. 
  Dashed line in bottom panel marks ${\cal L}={\cal L}_{\rm WF}$. 
  Inset shows the finite-$T$ conductivity (solid lines) $\sigma=GL/W$ 
  [see Eqs.\ (\ref{gland}) and (\ref{llndef}) in the main text] 
  and the linear fit (dash-dot line) to the corresponding $T=0$ dataset 
  in Fig.\ \ref{sigte0fig}. 
}
\end{figure}

In Fig.\ \ref{polylofig} we display the selected numerical data from Fig.\ 
\ref{seebwffig}, for $T=0.2\,$K and $T=2\,$K, as functions of $\mu/(k_BT)$
[solid lines] in order to compare them with predictions of the linear model 
for transmission-energy dependence $T(E)\propto{}|E|$ [dashed-dotted lines] 
elaborated in Appendix~A. 
For $T=2\,$K, both $S$ and ${\cal L}$ show an~agreement better than $10\%$
with the linear model for $\mu\lesssim{}E_L\approx{}5\,k_BT$. 
For $T=0.2\,$K, larger deviations appear for low chemical potentials due to 
the influence of transport via evanescent waves, which are significant for
$\mu<\hbar{}v_F/L\approx{}2-3\,k_BT$. 
For larger $\mu$, a~few-percent agreement with the linear model is restored, 
and sustained as long as $\mu\lesssim{}E_L\approx{}40\,k_BT$. 

Another remarkable feature of the results presented in Fig.\ \ref{polylofig}
becomes apparent when determining the extrema: The maximal thermopower 
corresponds to $\mu_{\rm max}^{(S)}/(k_BT)=2.0$ at $T=2\,$K, or to
$\mu_{\rm max}^{(S)}/(k_BT)=2.2$ at $T=0.2\,$K; the minimal Lorentz number 
corresponds to $\mu_{\rm min}^{({\cal L})}/(k_BT)=5.4$ at $T=2\,$K, or to
$\mu_{\rm min}^{({\cal L})}/(k_BT)=4.6$ at $T=0.2\,$K. 
In other words, an almost perfect agreement with the linear model 
[see, respectively, the second equality in Eq.\ (\ref{smaxexact}), 
or the second equality in Eq.\ (\ref{lorminexact}) in Appendix~A]  
is observed provided that
\begin{equation}
  \frac{\hbar{}v_F}{L}\ \ll{}\ 
  k_BT \sim \mu_{\rm max}^{(S)} \sim \mu_{\rm min}^{({\cal L})}
  \ \ll{}\ E_L. 
\end{equation}
In consequence, the effects that we describe may be observable for the sample
length $L>10\,\mu$m. 

%

\subsection{Electronic and phononic parts of the thermal conductance}

\begin{figure}
\centerline{
  \includegraphics[width=0.9\linewidth]{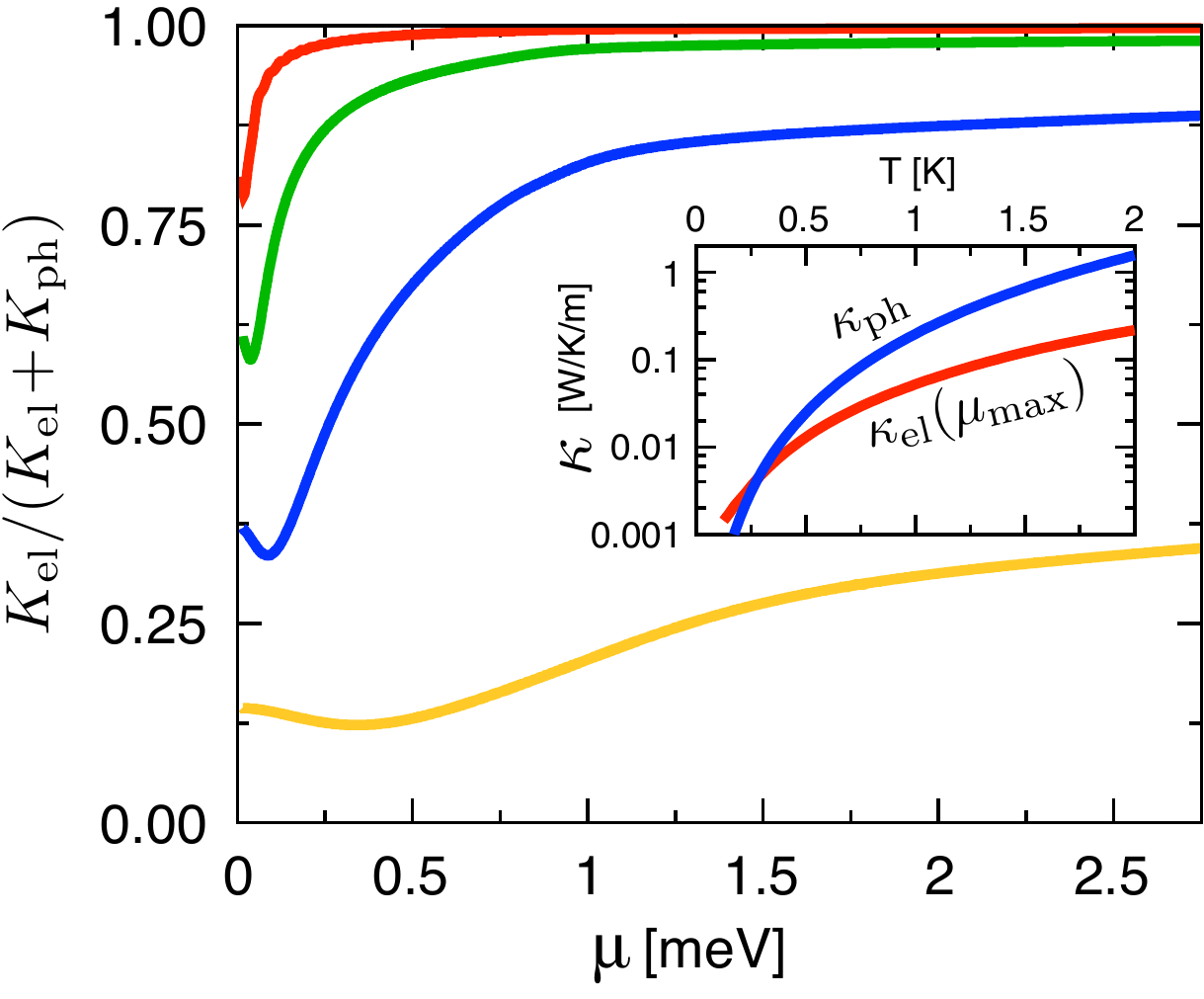}
}
\caption{ \label{kephfig}
  Relative electronic contribution to the thermal conductance for $t'=0.3\,$eV
  as a~function of the chemical potential. The temperatures are (top to bottom)
  $T=80\,$mK, $0.2\,$K, $0.515\,$K, and $2\,$K. 
  Inset shows phononic [blue line] and electronic [red line] thermal 
  conductivities ($\kappa=KL/(2dW)$, with $d=0.335\,$mn the 
  separation between layers) as functions of temperature, with the chemical
  potential fixed at $\mu=\mu_{\rm max}$ corresponding to the maximal Seebeck
  coefficient. 
}
\end{figure}

Before discussing the thermoelectric figure of merit $ZT$ we first display,
in Fig.\ \ref{kephfig}, values of the dimensionless prefactor in the last 
expression of Eq.\ (\ref{ztdef}), quantifying relative electronic 
contribution to the thermal conductance. 
The phononic transmission spectrum [see Eq.\ (\ref{kaphland})] was calculated 
numerically by employing, for the sample length $L=17.7\,\mu$m, 
the procedure presented by Alofi and Srivastava \cite{Alo13} adapting 
the Callaway theory \cite{Cal59} for mono-nand few-layer graphenes 
\cite{kaphfoo}.
The results show that in sub-Kelvin temperatures the electronic contribution 
usually prevails, even if the system is quite close to the charge-neutrality 
point, as one can expect for a~gapless conductor. For $T>1\,$K, however,
the phononic contribution overrules the electronic one in the full range
of chemical potential considered. 

A~direct comparison of the phononic and the electronic and thermal 
conductivities calculated in the physical units (see inset in 
Fig.\ \ref{kephfig}) further shows that, if the chemical potential is adjusted
to $\mu_{\rm max}\equiv{}\mu_{\rm max}^{(S)}$ for a~given temperature, both 
properties are of the same order of magnitude up to $T=2\,$K. 
Also, for $\mu=\mu_{\rm max}$, we find that 
$K_{\rm el}=K_{\rm ph}$ at the temperature $T_{\rm el-ph}\approx{}0.3\,$K, 
which is almost insensitive to the value of $t'$.

\subsection{Maximal performance versus temperature}

\begin{figure}
\centerline{
  \includegraphics[width=\linewidth]{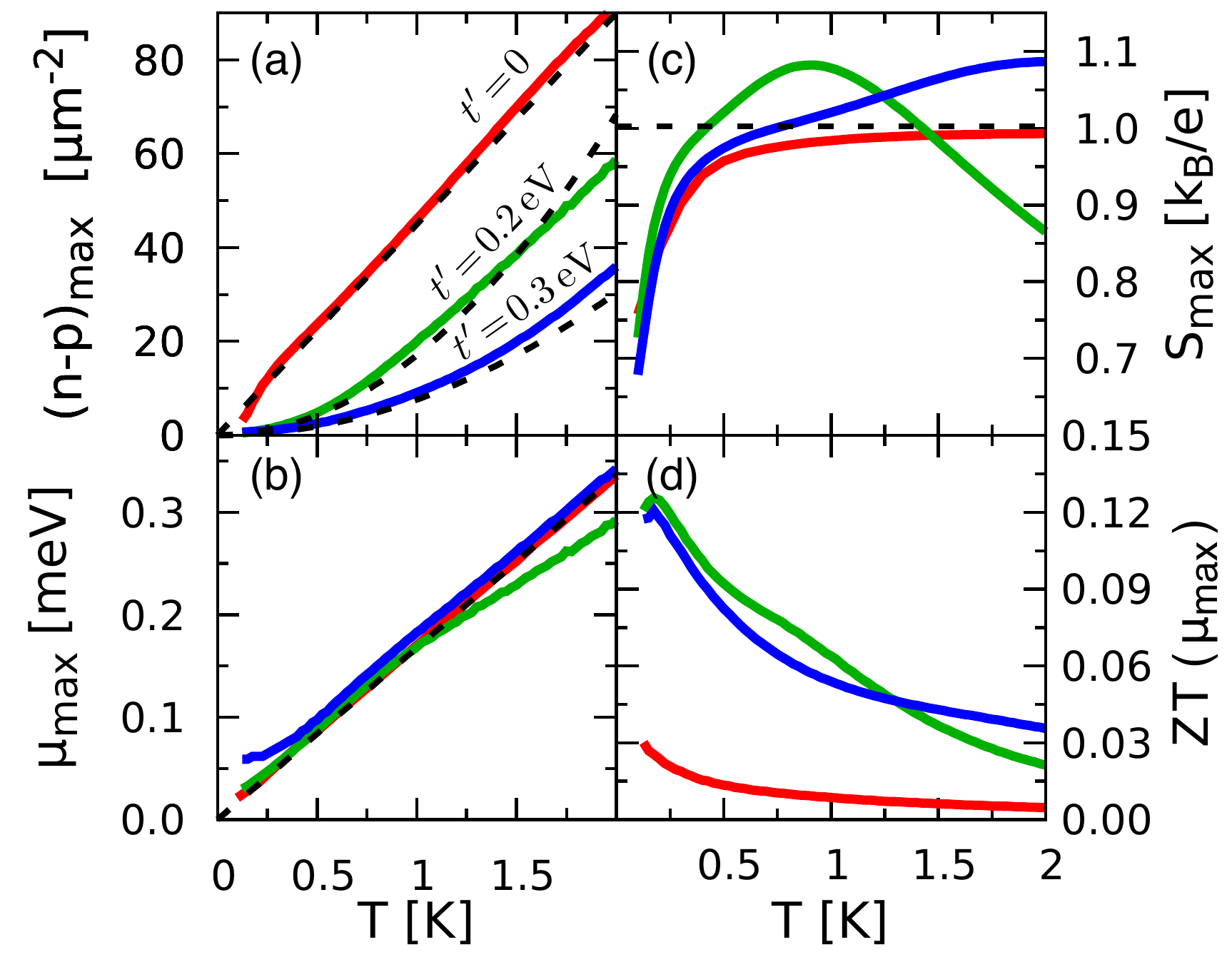}
}
\caption{ \label{smax4fig}
  Effective carrier concentration $(n\!-\!p)_{\rm max}$ (a) 
  and chemical potential $\mu_{\rm max}$ (b) 
  corresponding to the maximal Seebeck coefficient $S_{\rm max}$ (c)
  as functions of temperature. The figure of merit 
  $ZT(\mu_{\rm max})$ (d) is also displayed. Solid lines represent the numerical 
  results for different values of $t'$ [$\,$specified in panel (a)$\,$]. 
  Dashed lines mark predictions of the linear model for 
  transmission-energy dependence $T(E)\propto{}|E|$. 
}
\end{figure}

In Fig.\ \ref{smax4fig} we present parameters characterizing the maximal
thermoelectric performance for a~given temperature ($0<T\leqslant{}2\,$K).  
As the existing experimental works refer to the carrier concentration rather 
then to the corresponding chemical potential, we focus now on the functional 
dependence of the former on $T$ (and $t'$). 

Taking into account that the maximal performance is expected for 
$\mu\sim{}k_BT$ (see previous subsection), and that a~gapless system is 
under consideration, 
one cannot simply neglect the influence of minority carriers. 
For the conduction band ($\mu>0$), the {\em effective carrier concentration} 
can be written as 
\begin{align}
  n\!-\!p &= \int_0^{\infty} dE\,\rho(E)f(\mu,E) \nonumber\\ 
  &- \int_{-\infty}^0 dE\,\rho(E)[1-f(\mu,E)], 
  \label{enminuspe}
\end{align}
where we have supposed the particle-hole symmetry $\rho(E)=\rho(-E)$. 
[For the valence band ($\mu<0$), the effective concentration $p\!-\!n$ is 
simply given by the formula on the right-hand side of Eq.\ (\ref{enminuspe}) 
with an opposite sign.] 
Next, the approximating Eqs.\ (\ref{rhozero}) and (\ref{rhoapp}) for the 
density of states lead to 
\begin{align}
  n\!-\!p &\approx \rho_0{}k_BT \nonumber\\ 
  &\times
  \begin{cases} 
    { \displaystyle
      \,y, 
    } 
    & \ \text{if }\ t' = 0, \\
    & \vspace{-1em} \\
    {\displaystyle
      \,\tau_L\left[\,
        {\cal I}_1(y) + 0.33\,\tau_L^2\, {\cal I}_3(y)\,
      \right], 
    } 
    & \ \text{if }\ t' \neq 0,  
  \end{cases}
  \label{nmaxint}
\end{align}
where $y=\mu/k_BT$, $\tau_L=k_BT/E_L$, and we have defined 
\begin{equation}
  {\cal I}_n(y)=
  \int_{-y}^\infty{}\frac{(x+y)^n}{e^x+1}dx -
  \int_{y}^\infty{}\frac{(x-y)^n}{e^x+1}dx. 
\end{equation}
(In particular, ${\cal I}_0(y)=y$.) 
Numerical evaluation of the integrals in Eq.\ (\ref{nmaxint}) for 
$y=y_{\rm max}$ given by Eq.\ (\ref{smaxexact}) in Appendix~A brought us to
\begin{align}
  (n\!-\!p)_{\rm max} &\approx 
  \rho_0{}k_BT \nonumber\\
  &\times
  \begin{cases} 
    { \displaystyle
      1.949
    } 
    &  \text{if }\ t' = 0, \\
    & \vspace{-1em} \\
    {\displaystyle
      3.269\,\tau_L\!\left(1\!+\!3.23\,\tau_L^2\right)
    } 
    &  \text{if }\ t' \neq 0.  
  \end{cases}
  \label{nmaxlin}
\end{align}
In turn, the carrier concentration corresponding to the maximum of $S$
for a~given $T$ is determined by the value of $E_L$. 
[A~similar expression for the minimum of ${\cal L}$, see 
Eq.\ (\ref{lorminexact}) in Appendix~A, is omitted here.] 

Solid lines in Fig.\ \ref{smax4fig}(a) show the values of $(n\!-\!p)_{\rm max}$ 
calculated from Eq.\ (\ref{enminuspe}) for the actual density of states 
and the chemical potential $\mu=\mu_{\rm max}$ [displayed with solid lines in 
Fig.\ \ref{smax4fig}(b)] adjusted such that the Seebeck coefficient, 
obtained numerically from Eq.\ (\ref{sland}), reaches the {\em conditional} 
maximum ($S_{\rm max}$) [see Fig.\ \ref{smax4fig}(c)] at a~given 
temperature $T$ (and one of the selected values of $t'=0$, $0.2\,$eV, 
or $0.3\,$eV).  
The numerical results are compared with the linear-model predictions (dashed
lines in all panels), given explicitly by Eq.\ (\ref{nmaxlin}) 
[Fig.\ \ref{smax4fig}(a)] or
Eq.\ (\ref{smaxexact}) in Appendix~A [Figs.\ \ref{smax4fig}(b) and 
\ref{smax4fig}(c)]. 
Again, the linear model shows a~relatively good agreement with corresponding 
data obtained via the mode-matching method; 
moderate deviations are visible for $t'\neq{}0$ when 
$\mu_{\rm max}\gtrsim{}E_L/2$. 
In such a~range, both $\rho(E)$ no longer follows the approximating 
Eq.\ (\ref{rholin}), and the sudden rise of $T(E)$ near $E\approx{}E_L$ 
starts to affect thermoelectric properties. 

Fig.\ \ref{smax4fig}(c) and Fig.\ \ref{smax4fig}(d) display, respectively, 
the maximal Seebeck coefficient ($S_{\rm max}$) and figure of merit 
($ZT(\mu_{\rm max})$) as functions of temperature. 
For $t'\neq{}0$, the former shows broad peaks, centered near temperatures
corresponding to $\mu_{\rm max}\approx{}0.4\,E_L$, for which the prediction of 
the linear model [see Eq.\ (\ref{smaxexact}) in Appendix~A] 
is slightly exceeded (by less then $10\%$), whereas for
$t'=0$ a~monotonic tempereture dependence, approaching the linear-model value, 
is observed. The figure of merit (calculated for $\mu=\mu_{\rm max}$) shows
relatively fast temperature decay due to the role of phononic thermal
conductivity (see Sec.\ IIID). We find that $ZT(\mu_{\rm max})$, although
being relatively small, is noticeably elevated in the presence of trigonal 
warping in comparison to the $t'=0$ case. 

The behavior of $S_{\rm max}$ presented in Fig.\ \ref{smax4fig}(c)
suggests a~procedure, allowing one to determine the 
trigonal-warping strength via directly measurable quantities. 

\begin{figure}
\centerline{
  \includegraphics[width=0.9\linewidth]{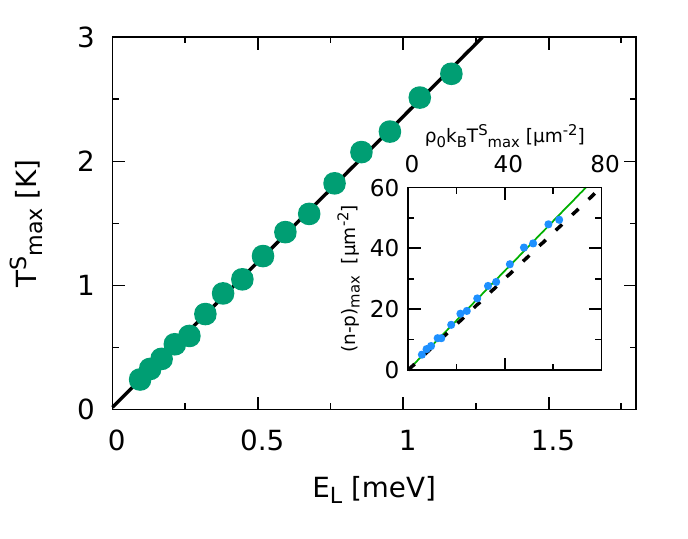}
}
\caption{ \label{tsmaxfig}
  Temperature corresponding to the maximal thermopower as a~function of
  the Lifshitz energy (datapoints). Least-squares fitted linear dependence 
  [see Eq.\ (\ref{tsmaxfit})] is also displayed (line). 
  Inset shows the carrier concentration versus 
  $\rho_0k_BT_{\rm max}^S$, together with the model prediction 
  (dashed line) and the linear fit (solid line) 
  [see Eqs.\ (\ref{nmaxlin}) and (\ref{nsmaxfit})]. 
}
\end{figure}

For any $t'\neq{}0$, one can determine a~unique {\em global} 
maximum of $S=S(\mu,T)$, which is reached at $\mu=\mu_{\rm max}^{S}$ and 
$T=T_{\rm max}^{S}$. 
Our numerical findings for $0.1\,\text{eV}\leqslant{}t'\leqslant{}0.35\,
\text{eV}$ are presented in Fig.\ \ref{tsmaxfig}, where we have
plotted (instead of $\mu_{\rm max}^{S}$), the optimal effective concentration 
$(n\!-\!p)_{\rm max}^{S}$ [see the inset].  
The best-fitted lines displayed Fig.\ \ref{tsmaxfig} are given by
\begin{align}
  T_{{\rm max},\,{\rm fit}}^S &= 0.203(2)\times E_L/k_B,  
  \label{tsmaxfit} \\
   \left(n-p\right)_{{\rm max},\,{\rm fit}}^S &= 0.816(5) \times 
   \rho_0{k_B}T_{\rm max}^S,
  \label{nsmaxfit} 
\end{align}
where numbers in parentheses are standard deviations for the last digit. 
A few-percent deviation of the actual $(n\!-\!p)_{{\rm max}}^S$ from predictions 
of the linear model [see dashed line in the inset, obtained from 
Eq.\ (\ref{nmaxlin}) by setting $\tau_L=k_BT_{{\rm max},\,{\rm fit}}^S/E_L
\approx{}0.20\,$] is relatively small taking into account that the existence
of a~global maximum of $S(\mu,T)$ is directly link to the breakdown of 
the linear model occurring for $\mu\sim{}k_BT\gtrsim{}E_L$ (and therefore 
is not observed in the $t'=0$ case). 

We further notice that 
Eqs.\ (\ref{tsmaxfit}) and (\ref{nsmaxfit}) provide direct relations between
the two independent driving parameters corresponding to the optimal 
thermopower, $T_{{\rm max}}^S$ and $(n\!-\!p)_{{\rm max}}^S$, and the trigonal 
warping strength quantified by $E_L$.

\section{Conclusions}
We have investigated the thermopower, violation of the Wiedemann-Franz law,
and the thermoelectric figure of merit, for large ballistic samples of
bilayer graphene in the absence of electrostatic bias between the layers
(a~gapless case) and close to the charge-neutrality point. 
Although the thermoelectric performance is not high in such a~parameter range, 
we find that low-temperature behavior of thermoelectric properties 
is determined by microscopic parameters of the tight-binding Hamiltonian, 
including the skew-interlayer hopping integral responsible for the trigonal 
warping, and by the relativistic nature of effective quasiparticles 
(manifesting itself in linear energy dependence of both the density of states 
and the electrical conductivity). 

In particular, at sub-Kelvin temperatures, clear signatures of the Lifshitz transition, having forms of anomalies in chemical-potential dependences of the Seebeck coefficient and the Lorentz number, occurs in a vicinity of the Lifshitz energy (defined by the microscopic parameters and quantifying the trigonal-warping strength). The anomalies are blurred out by thermal excitations above the crossover temperatures (different for the two thermoelectric properties) that are directly proportional to the Lifshitz energy. 

At higher temperatures (of the order of $1\,$K) the trigonal-warping strength 
can be determined from thermoelectric measurements following one of the two 
different approaches: 
(i) finding the carrier concentration corresponding to the maximal 
thermopower as a~function of temperature, or
(ii) finding the {\em optimal} temperature, i.e., such that thermopower 
reaches its global maximum. 
The first possibility is linked to the properties of massless quasiparticles, 
due to which the carrier concentration corresponding to the maximal thermopower
depends approximately quadratically on temperature and reciprocally on the 
Lifshitz energy.  
On the other hand, existence of unique optimal temperature 
(equal to $2\,K$ if the handbook value of the Lifshitz energy 
$E_L/k_B\approx{}10\,$K is supposed) 
is related the gradual conductivity enhancement, and subsequent 
suppression of the thermopower, with increasing population 
of thermally-excited massive quasiparticles above the Lifshitz energy. 

To conclude, we have show that thermoelectric measurements may complement 
the list of techniques allowing one to determine tight-binding parameters
of bilayer-graphene Hamiltonian. Unlike the well-established techniques 
\cite{Mac13} (or the other recently-proposed \cite{Rut14b,Rut16}), 
they neither require  high-magnetic field measurements nor refer to 
conductivity scaling with the system size. 
Instead, the proposed single-device thermoelectric measurements must be
performed on large ballistic samples (with the length exceeding $10\,\mu$m),
such that quantum-size effects define the energy scale much smaller then
the Lifshitz energy. 

As we have focused on clean ballistic systems, several factors which may 
modify thermoelectric properties of graphene-based devices, including the 
disorder \cite{Mac13}, lattice defects \cite{Dre10}, or magnetic impurities 
\cite{Uch11}, are beyond the scope of this study.  
However, recent progress in quantum-transport experiments on ultraclean 
free-standing monolayer samples exceeding $1\,\mu$m size \cite{Kum16,Ric15} 
allows us to expect that similar measurements would become possible in bilayer 
graphene soon. 
Also, as the effects we describe are predicted to appear away from the 
charge-neutrality point, the role of above-mentioned factors should be less
significant than for phenomena appearing precisely at the charge-neutrality
point, such as the minimal conductivity \cite{May11,Sam16}. 
Similar reasoning may apply to the role of interaction-induced 
spontaneous energy gap \cite{Rtt11,Fre12,Bao12} (we notice that experimental
values coincide with energy scales defined by quantum-size effects, 
e.g., $\hbar{}v_F/L\approx{}3\,$meV for $L=250\,$nm in Ref.\ \cite{Bao12}). 

{\em Note added in proof.\/} 
Recently,  we become aware of theoretical works on strained 
monolayer graphene reporting quite similar, double-peak spectra of 
the Seebeck coefficient for sufficiently high uniaxial strains \cite{Man17}.

\section*{Acknowledgments}
We thank to Colin Benjamin and Francesco Pellegrino for the correspondence, and to one of the Referees for pointing out the role the phononic part of the thermal conductance. 
The work was supported by the National Science Centre of Poland (NCN) via Grant No.\ 2014/14/E/ST3/00256. Computations were partly performed using the PL-Grid infrastructure. 
D.S.\ acknowledge the financial support from dotation KNOW from Krakowskie Konsorcjum ``Materia-Energia-Przysz{\l}o\'s\'c'' im.\ Mariana Smoluchowskiego.


\begin{widetext}
\appendix
\section{Linear model for transmission-energy dependence}

At sufficiently high temperatures, thermoelectric properties given by
Eqs.\ (\ref{gland}--\ref{ztdef}) become insensitive to the detailed 
functional form of $T(E)$, and simplified models can be considered. 
Here we assume  $T(E)={\cal C}|E|$, with ${\cal C}$ being a~dimensionless 
parameter. In turn, Eq.\ (\ref{llndef}) lead to 
\begin{align}
  L_0 &= \frac{{\cal D}}{\beta}
  \left( \ 
    y\int_0^y\frac{dx}{\cosh{x}+1} + \int_y^{\infty}\frac{x\,dx}{\cosh{x}+1}
    \ \right)
  = \frac{{\cal D}}{\beta} \ln\left(2\cosh{y}+2\right), 
  \label{l0exact}
  \\
L_1 &= \frac{{\cal D}}{\beta^2}
\left( \ 
  \int_0^y\frac{x^2dx}{\cosh{x}+1} + y\int_y^{\infty}\frac{x\,dx}{\cosh{x}+1}
\ \right) 
	= \frac{{\cal D}}{\beta^2}
	\left[ \ 
  	\frac{\pi^2}{3}+y^2-y\ln\left(2\cosh{y}+2\right)
  	+ 4\mbox{Li}_2(-e^{-y})
	\ \right], 
      \label{l1exact}
      \\
L_2 &= \frac{{\cal D}}{\beta^3}
\left( \ 
  y\int_0^y\frac{x^2dx}{\cosh{x}+1} + \int_y^{\infty}\frac{x^3dx}{\cosh{x}+1} 
\ \right)
 	\nonumber \\
	&= \frac{{\cal D}}{\beta^3}
	\left[ \ 
  	\frac{\pi^2}{3}y-y^3+y^2\ln\left(2\cosh{y}+2\right)
  	- 8y\mbox{Li}_2(-e^{-y}) - 12\mbox{Li}_3(-e^{-y}) 
	\ \right], 
      \label{l2exact}
\end{align}
where ${\cal D}=(g_sg_v/h)\,{\cal C}$, $\beta=1/k_BT$, $y=\beta\mu$, 
and $\mbox{Li}_s(z)$ is the polylogarithm function \cite{Old09}. 
Subsequently, the Seebeck coefficient and the Lorentz number 
(see Eqs.\ (\ref{sland}) and (\ref{lodef}) in the main text) are given by
\begin{align}
S &= \frac{k_B}{e}\, \left[\ 
  -y+\frac{\frac{\pi^2}{3}+y^2+4\mbox{Li}_2(-e^{-y})}{\ln\left(2\cosh{y}+2\right)}
\ \right], 
\label{seebvials} 
\\
\frac{K_{\rm el}}{TG} &= \left(\frac{k_B}{e}\right)^2\, \left\{\ 
  \frac{ {\pi^2}y+y^3-12\mbox{Li}_3(-e^{-y})}{\ln\left(2\cosh{y}+2\right)}
  - \left[\ 
  \frac{\frac{\pi^2}{3}+y^2+4\mbox{Li}_2(-e^{-y})}{\ln\left(2\cosh{y}+2\right)}
  \ \right]^2  
\ \right\}.   
\label{lorevials}
\end{align}

As the right-hand sides in Eqs.\ (\ref{seebvials}) and (\ref{lorevials})
depend only on a~single dimensionless variable ($y$) they are convenient
to be compared with thermoelectric properties obtained numerically via the 
mode-matching method (see Sec.\ III for details). 
In particular, 
the function of Eq.\ (\ref{seebvials}) is odd and has a~single maximum 
for $y>0$, i.e.\
\begin{equation}
  S_{\rm max}=1.0023\,k_B/e \ \ \ \ \text{for}\ \ \ \ y_{\rm max}^{(S)}=1.9488,
  \label{smaxexact}
\end{equation}
what is approximated by Eq.\ (\ref{smaxlimo}) in the main text. 
Analogously, the function of Eq.\ (\ref{lorevials}) is even, and has 
a~maximum at $y=0$, that brought us to Eq.\ (\ref{lormaxlimo}) in the main 
text. 
It also reaches a~minimum
\begin{equation}
  {\cal L}_{\rm min}=3.0060\,(k_B/e)^2\approx{}0.91\,{\cal L}_{\rm WF}
  \ \ \ \ \text{for}\ \ \ \ y_{\rm min}^{({\cal L})}=4.5895,
  \label{lorminexact}
\end{equation}
with the Wiedemann-Franz constant the 
${\cal L}_{\rm WF}=\frac{1}{3}\pi^2(k_B/e)^2$. 
For $y\rightarrow\infty$ we have ${\cal L}\rightarrow{\cal L}_{\rm WF}$. 

%
%

$\mbox{ }$

\end{widetext}




\begin{thebibliography}{}
\bibitem{elhofoo}
The combined particle-hole-reflection symmetry applies as we have neglected 
the next-nearest neighbor intralayer hopping $t_2$, which is not determined 
for BLG \cite{Mac13}. Effects of $t_2\sim{}t_\perp$ are insignificant when
discussing the band structure for $|E|\sim{}E_L\ll{}t_\perp$.
\bibitem{mcycfoo}
We notice a~generic relation between $\rho(E)$ and the cyclotronic mass 
$m_C(E)= (\pi{}\hbar^2/2)\rho(E)$.
\bibitem{theta0foo}
For $L=W/20=10^4\,l_\perp$, effects of the crystallographic orientation are 
negligible starting from  $|\mu|\gtrsim{}\hbar{}v_F/L\approx{}38\,\mu$eV, as
the summation over normal modes in Eq.\ (\ref{glandzero}) involves 
numerous contributions, corresponding to different angles of incident, that 
are comparable.
\bibitem{mottfoo}
Due to the presence of Fabry-P\'{e}rot oscillations, the well-known
Mott formula $S=(\pi^2/3)\,e^{-1}k_B^2T\times{}
\left[\,\partial\,{\ln{}T(E)}/\partial{}E\,\right]_{E=\mu}$ 
cannot be directly applied.
\bibitem{kaphfoo}
The model parameters are choosen such that the experimental results of
Ref.\ \cite{Pet11} are reproduced if $L=0.25\,\mu$m. 
Namely, the Callaway model parameters are
$B_U=4.77\times{}10^{-25}\,$s$\cdot{}$K$^{-3}$ and 
$B_N=3.18\times{}10^{-25}\,$s$\cdot{}$K$^{-3}$, 
the defect concentration is $c_d=1.6\times{}10^{-4}$, 
and the concentration of C$^{13}$ isotope is $10^{-2}$. 
Remaining parameters of the phonon dispersion relations are same as in
the first paper of Ref.\ \cite{Alo13}.
\end{thebibliography}
\end{document}